\documentclass[11pt]{article}

\usepackage{amsmath}  \usepackage{amsfonts}
\usepackage{graphicx,
xcolor }

\def\be{\begin{equation}}
\def\ee{\end{equation}}
\def\ba{\begin{eqnarray}}
\def\ea{\end{eqnarray}}
\def\nn{\nonumber}
\def\lb{\label}
\def\dfrac{\displaystyle\frac}
\def\bb{\bibitem}
\def\v{\overline{v}}
\def\E{{\cal E}}
\def\p{\hat\varphi}

\begin{document}
\begin{titlepage}
\date{}
\title{\begin{flushright}\begin{small}    LAPTH-016/17
\end{small} \end{flushright} \vspace{1.5cm}
A tale of two dyons}
\author{G\'erard Cl\'ement$^a$\thanks{Email: gclement@lapth.cnrs.fr},
Dmitri Gal'tsov$^{b,c}$\thanks{Email: galtsov@phys.msu.ru} \\ \\
$^a$ {\small LAPTh, Universit\'e Savoie Mont Blanc, CNRS, 9 chemin de Bellevue,} \\
{\small BP 110, F-74941 Annecy-le-Vieux cedex, France} \\
$^b$ {\small Department of Theoretical Physics, Faculty of Physics,}\\
{\small Moscow State University, 119899, Moscow, Russia }\\
$^c$ {\small  Kazan Federal University, 420008 Kazan, Russia}}

\maketitle

\begin{abstract}
We present a one-parameter family of stationary, asymptotically flat
solutions of the Einstein-Maxwell equations with only a mild
singularity, which are endowed with mass, angular momentum, a dipole
magnetic moment and a quadrupole electric moment. We briefly analyze
the structure of this solution, which we interpret as a system of
two extreme co-rotating black holes with equal masses and electric
charges, and opposite magnetic and gravimagnetic charges, held apart
by an electrically charged, magnetized string which also acts as a
Dirac-Misner string.
\end{abstract}
\end{titlepage}
\setcounter{page}{2}

\setcounter{equation}{0}
\section{Introduction}
Stationary asymptotically flat black holes with regular connected
event horizons are known to be strongly limited by uniqueness
theorems and given in the vacuum case by the Kerr metric, and in the
electrovacuum case by the Kerr-Newman solution. In the search for
alternatives to this standard black hole scenario, prompted by the
coming perspectives to receive high accuracy data from the center of
the Galaxy, one is led to relax some basic assumptions, among which
one could most safely sacrifice horizon connectedness. Many such
solutions of the Einstein-Maxwell equations known in the
Weyl-Papapetrou form are nicely reviewed in the Griffiths and
Podolsky book \cite{GP}. Leaving aside the well-known
Israel-Wilson-Perj\`es and Weyl linear superpositions, these
solutions can be roughly classified in two families. One was
obtained via soliton generating techniques and consists of
non-linear superpositions of aligned black holes possibly endowed
with charges and NUT parameters. The other contains solutions
generated by other methods, such as the static magnetized Bonnor
solution \cite{Bonnor66}, or the one-parameter class of static
Zipoy-Voorhees (ZV) \cite{Z,V} vacuum solutions, also known as
$\gamma$-metrics, and their rotating Tomimatsu-Sato (TS) \cite{TS}
cousins with integer $\gamma$, the $\gamma=1$ TS solution coinciding
with the Kerr metric. The physical interpretation of these last
solutions is far from trivial, and it was only relatively recently
recognized that the Bonnor solution \cite{Emparan} or the TS
solution with $\gamma=2$ (TS2) \cite{KoHi} actually describe
black-hole pairs. Both families generically contain naked ring-type
curvature singularities, and/or milder conical line singularities
(strings), which however can be avoided by imposing external fields
\cite{Emparan}, at the expense of asymptotic flatness. Here we adopt
a tolerant attitude with respect to mild naked singularities and do
not reject solutions endowed with novel and very intriguing features
for the sake of cosmic censorship. It is worth noting that the
possibility of violation of cosmic censorship in gravitational
collapse has received more attention recently \cite{Yagi}.

In this short note (a more detailed version will be published
elsewhere) we wish to draw attention to a new electrovacuum solution
obtained some time ago by one of us \cite{GC98} via an original
generating technique which produces a one-parameter family of
rotating electrovacuum solutions from a given static vacuum one.
Application of this procedure to the static ZV metric generates a
rotating solution with a magnetic dipole moment, which is the unique
so far known rotating generalization of the ZV metric with
non-integer $\gamma$, and has the advantage to be free from a naked
ring singularity. For $\gamma=1$ it is again the Kerr metric, while
for other integer values of $\gamma$ it is not the vacuum TS metric,
but some new electrovacuum metric. Postponing the case of generic
real $\gamma$ for future publication, we concentrate here on the
$\gamma=2$ version which looks the most interesting physically.

\setcounter{equation}{0}
\section{The solution}
This one-parameter family of rotating solutions was generated from
the static $\gamma=2$ Zipoy-Voorhees (ZV2) vacuum solution in
\cite{GC98}. The Ernst potentials of the rotating solution are
${\cal E} = (U-W)/(U+W)$, $\psi = V/(U+W)$, with the Kinnersley
potentials:
 \ba\lb{kin}
U &=& p\dfrac{x^2+1}{2x} + iqy, \qquad V=\varepsilon(W-1),\nn\\
W &=& 1 + \dfrac{q^2}2\dfrac{1-y^2}{x^2-1} - i\dfrac{pq}2\dfrac{y}x, \qquad (\varepsilon^2=1)
 \ea
where the prolate spheroidal coordinates $x\ge1$, $y\in[-1,+1]$ are related to
the Weyl cylindrical coordinates $\rho$, $z$ by
 \be\lb{sphero}
\rho = \kappa\sqrt{(x^2-1)(1-y^2)}, \quad z = \kappa xy,
 \ee
(the positive constant $\kappa$ setting the length scale), and the
real parameters $p$ and $q$ are related by $p=\sqrt{1-q^2}$. The
potentials of the ZV2 solutions are recovered for $q=0$.

The form (\ref{kin}) of the solution is only implicit. Dualization of the imaginary
part of the scalar Ernst potentials to vector potentials leads to the explicit
metric and electromagnetic fields
 \ba\lb{ansatz}
ds^2 &=& - \frac{f}\Sigma\left(dt-\omega
d\varphi\right)^2 +
\kappa^2\Sigma\left[e^{2\nu}\left(\frac{dx^2}{x^2-1}
+ \frac{dy^2}{1-y^2}\right)\right. \nn\\
&& \left. \quad + f^{-1}(x^2-1)(1-y^2)d\varphi^2\right], \nn\\
A &=& \frac{\varepsilon}{\Sigma}[\v dt + \kappa\Theta d\varphi],
 \ea
($\varphi$ periodic with period $2\pi$), the various functions
of $x$ and $y$ occurring in (\ref{ansatz}) being
 \ba\lb{rotgam2}
f &=& \frac{p^2(x^2-1)^2}{4x^2} - \frac{q^2x^2(1-y^2)}{x^2-1}, \nn\\
\Sigma &=& \left[\frac{px^2+2x+p}{2x} +
\frac{q^2(1-y^2)}{2(x^2-1)}\right]^2 +
q^2\left(1-\frac{p}{2x}\right)^2y^2, \nn\\
e^{2\nu} &=& \frac{4x^2(x^2-1)^2}{p^2(x^2-y^2)^3},\qquad \omega= \frac{\kappa\Pi}{f}, \nn\\
\Pi &=& \Pi_1(x)(1-y^2) + \Pi_2(x)(1-y^2)^2,\nn\\
\Pi_1 &=& -\frac{q}{2p}\left\{\frac{(px+2)[4x^2+p^2(x^2-1)]}{x^2} +
\frac{4p(1+p^2)x+8 + p^2  - p^4}{x^2-1}\right\}, \nn\\
\Pi_2 &=& -\frac{q^3}2\left[\frac{p}{4x^2} +
\frac{2x-p}{x^2-1}\right], \nn\\
\v &=& \frac{q^2}4\left\{-\frac{p(2x-p)}{x^2} +
\left[\frac{p(2x-p)}{x^2} + \frac{px^2+2x+p}{x(x^2-1)}\right](1-y^2)
+ \frac{q^2(1-y^2)^2}{(x^2-1)^2} \right\}, \nn\\
\Theta &=& \Theta_1(x)(1-y^2) + \Theta_2(x)(1-y^2)^2 + \Theta_3(x)(1-y^2)^3,\nn\\
\Theta_1 &=& -\frac{q}{4p} \left\{\frac{p\left[p^2x^3 + 5px^2 -
(8-4p^2+p^4)x + 2p-3p^3\right]}{x^2}
+ \frac{(16-p^2+p^4)px + 8+7p^2+p^4}{x^2-1}\right\},\nn \\
\Theta_2 &=&
-\frac{q^3}{8p}\left[\frac{p(4x^3-3px^2+2p^2x+5p)}{x^2(x^2-1)}
+ \frac{2(4x+3p+p^3)}{x(x^2-1)^2}\right],\nn \\
\Theta_3 &=& -\frac{q^5}{8x(x^2-1)^2}.
 \ea

The metric (\ref{ansatz}),(\ref{rotgam2}) is
asymptotically (for $x\to\infty$) Minkowskian in spherical
coordinates, with $r = \kappa x$, $\cos\theta=y$. The associated
conserved charges are the mass $M$, angular momentum $J$, dipole
magnetic moment $\mu$, and quadrupole electric moment $Q_2$:
 \be\lb{aschar}
M = \frac{2\kappa}p, \quad J = \frac{\kappa^2 q(4+p^2)}{p^2},
\quad \mu = \varepsilon\kappa^2q, \quad Q_2 = -\varepsilon\kappa^3q^2/p.
 \ee
The ratio $|\mu/J|$ is bounded above by $1/5$, in agreement with the
Barrow-Gibbons bound \cite{BG} $|\mu/J|\le1$, while the ratio $|J|/M^2$
satisfies the Kerr-like bound
 \be
|J|/M^2 \le 1,
 \ee
the upper bound being attained in the limit $p\to0$ with $M$ fixed, in
which case the solution reduces to the extreme Kerr vacuum solution.

To the difference of the TS2 vacuum solution, which is another
rotating generalization of the ZV2 solution, the present solution is
free from a naked ring singularity. Such a singularity would arise
as a zero of the function $\Sigma(x,y)$, however it is clear from
(\ref{rotgam2}) that $\Sigma$ admits the lower bound
$\Sigma(x,y)>(p+1)^2$. Another possible locus for singularities is
the symmetry axis $\rho=0$ which from (\ref{sphero}) can be divided
in three pieces: $z>\kappa$ ($y=+1$, $x>1$), $-\kappa<z<\kappa$
($x=1$, $-1<y<1$), and $z<-\kappa$ ($y=-1$, $x>1$). By construction
the solution is regular for $y=\pm1$ if $x\neq1$. However it
presents coordinate singularities on the segment $S$
($-\kappa<z<\kappa$) as well as at its two ends $H_\pm$
($z=\pm\kappa$). Other coordinate singularities are the {\em
ergosurface} $F \equiv f/\Sigma = 0$, which consists of two disjoint
components: the surface $f(x,y) = 0$, which contains $S$ ($f<0$ for
$x^2\to1$), and $S$ itself, on which $\Sigma\to\infty$; the {\em
causal boundary}, where $g_{\varphi\varphi} \equiv F^{-1}\rho^2 -
F\omega^ 2 = 0$, which contains $S$ ($F^{-1}\rho^2<0$ and $\omega$
finite) and is contained within the ergosurface component $f=0$; and
the {\em horizons}, where $N^2 \equiv \rho^2/g_{\varphi\varphi}=0$
with $g_{\varphi\varphi}>0$, obvious candidates being the the end
points $H_\pm$ of the segment $S$.

A first integral of the  geodesic equation in the metric (\ref{ansatz}) is
$ds^2/d\lambda^2=\epsilon$ (with $\lambda$ an affine parameter, and
$\epsilon=-1,0,1$), which can be written as
 \be
T+U=\epsilon,
 \ee
with
 \ba
T &=& \kappa^2\Sigma e^{2\nu}\left(\dfrac{\dot{x}^2}{x^2-1}
+ \dfrac{\dot{y}^2}{1-y^2}\right) > 0,\nn\\
U &=& \dfrac{(L-E\omega)^2F}{\rho^2} - \dfrac{E^2}F,
 \ea
where $\dot{} = d/d\lambda$, and $E$ and $L$ are the conserved
energy and orbital angular momentum. Near the segment $S$, where
$\xi^2 \equiv x^2-1\to0$, $-F$ and $\rho^2$ both scale as $\xi^2$,
so that for $E\neq0$ geodesics turn back before reaching $S$, while
geodesics with $E=0$ terminate on $S$, but cannot originate from
$\infty$ if they are timelike or null ($\epsilon= -1$ or $0$), so
that $S$ is a ``harmless'' naked singularity. Near the two endpoints
$H_\pm$, geodesics such that, near $x=1$, $1-y^2 \sim X^2(x^2-1)$
with $X$ fixed can be continued through $x=\pm y=1$ to a region with
$x<1$ and $y^2>1$, suggesting that these are actually two double
horizons. We thus arrive at a possible interpretation of this
solution as describing a system of two black holes $H_{\pm}$ held
apart by a string $S$, which shall be validated in the following.

\setcounter{equation}{0}
\section{The horizons}
When $x$ decreases from infinity to 1, an ergosurface $f(x,y)=0$ appears.
As $x$ goes to 1, $f$ goes to $-\infty$, unless $y$ goes simultaneously
to 1 with the ratio
 \be\lb{X2}
X^2 = \frac{1-y^2}{x^2-1}
 \ee
held fixed. Then,
 \ba\lb{fhor}
f &\to& -q^2X^2, \quad \Pi \to -qX^2\lambda(p),\nn\\
\Sigma &\to& \frac{p\lambda(p)}2 + q^2(1+p)X^2 + \frac{q^4}4X^4,
 \ea
with
 \be
\lambda(p) = \frac{(1+p)(8-4p+5p^2-p^3)}{2p} \ge 8
 \ee
(the lower bound being attained in the limit $q\to0$).
Rewriting the metric in the ADM form as
 \be
ds^2 = - N^2\,dt^2 + g_{ij}(dx^i+N^i\,dt)(dx^j+N^j\,dt),
 \ee
we then see that
 \be\lb{gpp}
g_{\varphi\varphi} = \kappa^2\left[\frac{\Sigma}f(x^2-1)(1-y^2) -
\frac{\Pi^2}{\Sigma f}\right]
 \ee
goes over to a positive function of $X$, while the lapse function
 \be
N^2 = \frac{\kappa^2(x^2-1)(1-y^2)}{g_{\varphi\varphi}}
 \ee
develops a double zero at $x=1$, $y=\pm1$, corresponding to two double horizons,
co-rotating at the angular velocity
 \be\lb{OmH}
\Omega_H = - N^\varphi\vert_H = \left. - \frac{\kappa\Pi}{\Sigma
g_{\varphi\varphi}}\right\vert_H = \left.
\frac{f}{\kappa\Pi}\right\vert_H = \frac{q}{\kappa\lambda(p)}.
 \ee

To investigate the geometry of the horizons, we follow \cite{KoHi}
and transform from the prolate spheroidal coordinates to the
coordinates $X$ (the positive square root of (\ref{X2})) and $Y =
y/x$. On the horizons $Y = \pm1$, the metric degenerates to
 \be\lb{methor}
ds_H^2  = \frac{4\kappa^2\Sigma(X)dX^2}{p^2(X^2+1)^4} +
\frac{\kappa^2\lambda^2(p)X^2d\p^2}{\Sigma(X)},
 \ee
in the co-rotating near-horizon frame $(\hat{t},X,Y,\p)$ defined by
$\hat{t} = t$, $\p = \varphi - \Omega_h t$. Introducing then a new
angular coordinate $\eta$ by
 \be\lb{eta}
X = \tan(\eta/2) \quad (0 \le \eta \le \pi),
 \ee
(\ref{methor}) can be rewritten as
 \be\lb{methor1}
ds_H^2 = \frac{\kappa^2\lambda(p)}{2pl(\eta)}\left(d\eta^2 +
l^2(\eta)\sin^2\eta d\p^2\right),
 \ee
where
 \be
l(\eta) = \frac{p\lambda(p)}2\frac{(X^2+1)^2}{\Sigma(X)}
 \ee
is everywhere positive and finite, and such that $l(0) = 1$. It
follows that each horizon is homeomorphic to $S^2$, with an area
 \be
{\cal A} = 4\pi\kappa^2\frac{\lambda(p)}{2p}.
 \ee
The corresponding areal radius is of the order of the total mass
$M$. At $\eta=\pi$ ($X\to\infty$), the metric (\ref{methor1})
presents a conical singularity with deficit angle $2\pi(1-\alpha)$,
where
 \be\lb{alpha}
\alpha = l(\pi) = \frac{2p\lambda(p)}{q^4} > \frac8{q^4} > 8.
 \ee

The evaluation of the electromagnetic potential on the horizon leads
to
 \be\lb{Ah}
\hat{A}_H= -\varepsilon\left(\frac{q^2(2-p)}{2\lambda(p)}\,dt +
\frac{\kappa q}4 \frac{(\delta(p)X^2+q^2\gamma(p)X^4)}{\Sigma(X)}\,
d\p\right)
 \ee
in the co-rotating frame, with
 \be
\gamma(p) = \frac{(1 + p)(4 - p + p^2)}p, \quad \delta(p) =
\frac{(1+p)^2(8-p^2+p^3)}p.
 \ee
Using the Tomimatsu formula \cite{tom84}
 \be
Q_H = -\dfrac1{4\pi}\oint_H\omega\,d\,{\rm Im}\psi\,d\varphi,
 \ee
with $\omega_H=1/\Omega_H$, we find that the horizons carry electric charges
 \be\lb{QH}
Q_+ = Q_- = -\dfrac{\varepsilon\kappa(1+p)}2,
 \ee
which also means that, to ensure global electric neutrality, the
string must be also charged, which we shall check in the next
section. The vector potential (\ref{Ah}) generates a magnetic field
perpendicular to the horizon. Because the normals to the two
horizons $Y = 1$ and $Y = -1$ are oppositely oriented and the net
magnetic charge is zero, the magnetic lines of force must emerge
from one horizon and flow into the other horizon, so that the two
horizons can be considered as carrying exactly opposite magnetic
charges
 \be\lb{P}
P_+=-P_- = \frac1{4\pi}\oint_{H_+}dA_{\varphi}\,d\varphi =
\varepsilon\frac{\kappa\gamma(p)}{2q}.
 \ee

Using the Ostrogradsky theorem and the Einstein-Maxwell equations,
the Komar mass and angular momentum at infinity
 \be
M = \frac1{4\pi}\oint_\infty D^\nu k^{\mu}d\Sigma_{\mu\nu}, \quad J
= -(1/8\pi)\oint_\infty D^\nu m^{\mu}d\Sigma_{\mu\nu}
 \ee
($k^\mu = \delta^\mu_t$, $m^\mu = \delta^\mu_\varphi$) can be
transformed \cite{tom84} into the sums over the boundary surfaces
(here, the two horizons and the string) $M = \sum_n M_n$, $J =
\sum_n J_n$, with
 \ba\lb{MJn}
M_n &=&
\frac1{8\pi}\oint_{\Sigma_n}\left[g^{ij}g^{ta}\partial_jg_{ta}
+2(A_t F^{it}-A_\varphi F^{i\varphi})\right]d\Sigma_i. \nn\\
J_n &=& -\frac1{16\pi}\oint_{\Sigma_n}\left[g^{ij}g^{ta}
\partial_jg_{\varphi a} +4A_\varphi F^{it}\right]d\Sigma_i.
 \ea
As shown by Tomimatsu, these reduce on the horizons to
 \ba
M_H &=& \frac1{8\pi}\oint_H\left[\omega\,d\,{\rm Im}\E +
2d(A_\varphi\,{\rm Im}\psi)\right]d\varphi,\nn\\
J_H &=& \frac1{8\pi}\oint_H\omega\left[\dfrac12\omega\,d\,{\rm Im}\E
+ d(A_\varphi\,{\rm Im}\psi) + \omega\hat{A}_t\,d\,{\rm
Im}\psi\right]d\varphi,
 \ea
which yield
 \ba
&M_+ = M_- =& \frac\kappa{p} +\frac{\kappa p}2,\lb{MH}\\
&J_+ = J_- =&
\frac{\kappa^2}{8qp}\left[2\lambda(p)(2+p^2)+q^2p(1+p)(2-p)\right].\lb{JH}
 \ea
The horizon mass (\ref{MH}) is larger than half of the global mass
$2\kappa/p$, so the string must have negative mass. Finally, we will
show in the next section that the two horizons carry opposite NUT
charges
 \be\lb{N}
-N_+=N_- = \dfrac{\kappa\lambda(p)}{4q}.
 \ee

\setcounter{equation}{0}
\section{The string}
For $\xi^2 \equiv x^2-1\to0$ (with $y^2<1$), the solution
(\ref{ansatz}) reduces to:
 \ba
ds^2 &\sim& - \frac{\kappa^2q^2}4(1-y^2)^2\,d\varphi^2 +
\frac{\kappa^2q^4}{p^2(1-y^2)}\left[\frac{dy^2}{1-y^2}\right. \nn\\
&& \left. + d\xi^2 + \alpha^2\Omega_H^2\xi^2\left(dt -
\kappa\left(\frac{\lambda(p)}q+q(1-p/2)(1-y^2)\right)
d\varphi\right)^2\right] \lb{rodmet}\\
A &\sim&
\varepsilon\left[\left(1-\frac{2(1+p)\xi^2}{q^2(1-y^2)}\right)dt -
\kappa\left(\frac{\gamma(p)}{q}
+\frac{q(1-y^2)}2\right)d\varphi\right], \lb{rodA}
 \ea
where we have neglected irrelevant terms of order $\xi^2$ and
higher. The singularity at $\xi=0$ is obviously a conical one, with
finite Ricci square scalar
 \be
R^{\mu\nu}R_{\mu\nu} \sim \frac{64p^4}{\kappa^4q^{12}}[(1+p)^2 +
q^2y^2]^2.
 \ee
Transforming to the horizon co-rotating frame by $d\varphi = d\p +
\Omega_h\,dt$, we find that the near-string metric (\ref{rodmet})
transforms to that of a spinning cosmic string \cite{DJH,GC85} in a
curved spacetime,
 \ba
ds^2 &\sim&
q^4\left[-\frac{(1-y^2)^2}{4\lambda^2(p)}\left(dt+\Omega_H^{-1}
\,d\p\right)^2\right. \nn\\
&+& \left.\frac{\kappa^2}{p^2(1-y^2)}\left(\frac{dy^2}{1-y^2} +
{d\xi^2} + \alpha^2\xi^2\,d\p^2\right)\right],
 \ea
with (negative) tension per unit length $(1-\alpha)/4$  where
$\alpha$ is given in (\ref{alpha}) , and ``spin'' $\Omega_H^{-1}/4$
 where $\Omega_H$ is given in (\ref{OmH}).

In view of the fact that the finite-length string connects two black
holes, this spin should actually be interpreted as a gravimagnetic
flow along the Misner string connecting two opposite NUT sources at
$\rho=0$, $z = \pm\kappa$, with the gravimagnetic potential $\omega/2 =
N_+\cos\theta_+ + N_-\cos\theta_-$ where $\cos\theta_\pm=\mp1$   along
the string, with $N_\pm$ given by (\ref{N}). Similarly, the constant
contribution $-\varepsilon\kappa\gamma(p)/q$ to $A_\varphi$ in
(\ref{rodA}) should be interpreted as the magnetic flow along a
Dirac string connecting two opposite monopoles at $z=\pm\kappa$ with
magnetic charges given by (\ref{P}). The non-constant contribution
gives rise to a magnetic field density $\sqrt{|g|}B^\xi =
F_{y\varphi} = \varepsilon\kappa qy$, which leads to an intrinsic
string magnetic moment
 \be
\mu_S = \frac1{4\pi}\int_{-1}^{+1}\sqrt{|g|}B^\xi\,z\,2\pi\,dy =
\frac{\varepsilon\kappa^2q}3 = \frac\mu3
 \ee
(to obtain the total magnetic moment $\mu$, the magnetic dipole
contribution $2\kappa P_H$ and the sum of the horizon magnetic
moments must be added to this).

Although it is not immediately obvious from (\ref{rodA}), the string
also carries an electric charge. The near-string covariant component
$F_{t\xi}$ of the radial electric field vanishes to order ${\rm
O}(\xi)$, but on account of $g_{tt} = {\rm O}(\xi^2)$ and
$\sqrt{|g|} = {\rm O}(\xi)$, the radial electric field density
$\sqrt{|g|}F^{t\xi}$ is finite and constant along the string,
leading to the electric charge
 \be
Q_S = \frac1{4\pi}\int_{-1}^{+1}\sqrt{|g|}F^{t\xi}\,2\pi\,dy =
\varepsilon\kappa(1+p).
 \ee
This string electric charge together with the horizon electric
charges lead to a vanishing total electric charge
 \be
Q_++Q_-+Q_S=0,
 \ee
a vanishing electric dipole moment, and a contribution to the total
electric quadrupole moment, to which must be added that of the two
opposite horizon electric dipole moments generated by the rotation
of the horizon magnetic charges, and the sum of the horizon electric
quadrupole moments.

The string mass and angular momentum can be evaluated from
(\ref{MJn}) integrated over the string $x=1$, $-1<y<1$, and are the
sum of gravitational and electromagnetic contributions. Although in
the co-rotating frame the string is a spinning cosmic string with
negative tension, and thus presumably negative gravitational mass,
in the global frame the gravitational contribution to the string
mass is -- surprisingly -- positive. However it is overwhelmed by
the negative electromagnetic contribution $-Q_SA_t(\xi=0)$,
resulting in a net negative string mass
 \be
M_S = \kappa - \kappa(1+p) = - \kappa p,
 \ee
which represents the binding energy between the two black holes of
mass $(\kappa/p + \kappa p/2)$, leading to the total mass
 \be
M = M_++M_-+M_S = \frac{2\kappa}p.
 \ee
The fact that the string mass is negative explains the repulsion
experienced by test particles in geodesic motion near the string
(antigravity).

Similarly, the string angular momentum is the sum of gravitational
and electromagnetic contributions
 \ba\lb{JS}
J_S &=&
\kappa^2\left[\frac{\lambda(p)}{2q}+\frac{q}3\left(1-\frac{p}2
\right)\right] -
\kappa^2(1+p)\left[\frac{\gamma(p)}{q}+\frac{q}3\right]\nn\\
&=& \frac{\kappa^2}{2q}\left[\lambda(p) - 2(1+p)\gamma(p) -
pq^2\right].
 \ea
The first term $\kappa^2\lambda(p)/2q$ is the NUT dipole $2\kappa
N_H$, the remainder corresponding to the intrinsic string angular
momentum. It can be checked that the horizon angular momenta
(\ref{JH}) and the total string angular momentum (\ref{JS}) add up
to the net angular momentum (\ref{aschar}):
 \be
J = J_+ + J_- + J_S = \frac{\kappa^2 q(4+p^2)}{p^2}.
 \ee

\setcounter{equation}{0}
\section{Summary and  outlook}

Summarizing the features of the solution, we can present it
schematically as shown in Fig.1. It is a system of two extreme
co-rotating black holes endowed with masses, NUT charges, electric
and magnetic charges held apart by a magnetized, electrically
charged string of negative tension, which is also a Dirac and Misner
string.
\begin{figure}[tb]
\begin{center}
\begin{minipage}[t]{0.48\linewidth}
\hbox to\linewidth{\hss%
  \includegraphics[width=1.5\linewidth,height=1.2\linewidth]{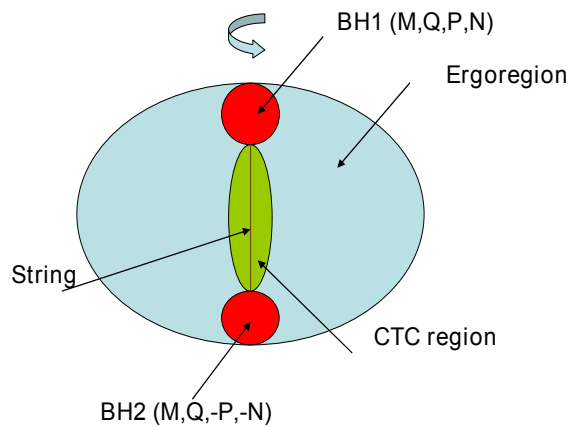}
\hss} \caption{\small  Schematic picture of the solution.} \label{F7}
\end{minipage}
\end{center}
\end{figure}
The whole system lies inside the ergosphere. As the Misner string
does not extend to infinity, the solution is asymptotically flat.
There are no strong curvature singularities, while the string, which
can formally be considered as a mild naked singularity, is
inaccessible from infinity. The charges compensate each other so
that the asymptotic parameters are the mass, the angular momentum
and the gravitational and electromagnetic multipole moments. The
family interpolates between the static vacuum ZV2 solution for $q=0$
and extreme Kerr for $q=1$. As usual, the Misner string is
surrounded by a region containing CTCs. We have
argued elsewhere \cite{RTN,NW} that NUT-related CTCs do not
necessarily lead to observable violations of causality.

As will be discussed elsewhere \cite{CGprep}, this solution can be
analytically continued beyond the horizons. The most economical
maximal analytical extension contains two interior regions between
an outer and an inner horizon (both degenerate), and beyond the
inner horizons a third region extending to spacelike infinity and
containing a timelike ring singularity.

Comparing with other known stationary solutions describing two-black
hole systems, we think that this one has a minimal number of
physically undesirable features and can be considered as ``almost''
physical. This surprising property is presumably related to the
specific generating technique of \cite{GC98} which endows the static
seed solution with rotation and charges, all depending on a single
parameter. Applying this transformation to the general-$\gamma$ ZV
solution generically gives singular space-times with novel features
\cite{CGprep}, which may be interesting in view of recent
discussions of alternatives to Kerr in astrophysical modeling
\cite{Yagi}. It is worth noting that our solution belongs to a
subclass of the nine-parameter family of solutions of the
Einstein-Maxwell equations constructed in \cite{manko00}, whose
physical features are still unexplored.

\section*{Acknowledgments} DG thanks LAPTh Annecy-le-Vieux for hospitality at different stages of
this work. DG also acknowledges the support of the Russian
Foundation of Fundamental Research under the project 17-02-01299a
and the Russian Government Program of Competitive Growth of the
Kazan Federal University.

\end{document}